\documentclass[
    ,final            
  ]
  {aipproc}

\layoutstyle{8x11single}

\usepackage{bm,amsmath}
\begin{document}

\title{Total electronic Raman scattering in the
charge-density-wave phase of the spinless Falicov-Kimball model}

\classification{71.10.Fd, 71.45.Lr, 78.30.-j} \keywords {Dynamical
mean-field theory, electronic Raman scattering, charge-density-wave phase}

\author{O.~P.~Matveev}{
  address={Institute for Condensed Matter Physics of the National Academy
of Sciences of Ukraine, \\
1 Svientsitskii Str., 79011 Lviv, Ukraine} }

\author{A.~M.~Shvaika}{
  address={Institute for Condensed Matter Physics of the National Academy
of Sciences of Ukraine, \\
1 Svientsitskii Str., 79011 Lviv, Ukraine} }

\author{J.~K.~Freericks}{
  address={Department of Physics, Georgetown University, Washington,
DC 20057, USA}
 }

\begin{abstract}

The total electronic Raman scattering spectrum, including the nonresonant, mixed and
resonant components, is determined for the charge-density-wave
(CDW) phase of the spinless Falicov-Kimball model at half filling
within dynamical mean-field theory. Its frequency dependence is
investigated for different values of the energy of the incident
photons. The spectra reflect the different structures in the density of
states and how they are modified by screening and resonance effects. The calculations are performed for the $B_{\rm 1g}$, $B_{\rm 2g}$
and $A_{\rm 1g}$ symmetries (which are typically examined in
experiment). Our results for the resonance effects
of the Raman spectra, found by tuning the energy of the incident photons, give information
about the many-body charge dynamics of the CDW-ordered phase.
\end{abstract}

\maketitle


\section{Introduction}
Experiments on inelastic light scattering are employed to
learn about the complicated charge dynamics of a wide class of strongly correlated electronic materials. The photon
couples to the charge excitations during the inelastic scattering process and directly probes the charge excitations of different symmetries.  In this work, we study the strongly correlated electron systems with
charge-density-wave (CDW) ordering. CDW systems possess a static
spatial modulation of the electronic charge with some  spatial ordering wavevector. Since the
underlying ionic cores are charged, they also respond to this
charge modulation of the electron density and often create a distorted
lattice structure that follows the modulated charge order of the
electrons.  A direct measurement of the lattice distortion due
to the ionic displacement is often the best way to measure the presence of CDW order; it is more difficult to directly
measure the electronic charge modulation in the material.

In the present work, we investigate how CDW order affects
inelastic light scattering experiments by utilizing
dynamical mean-field theory (DMFT) to exactly solve for the total electronic Raman spectra. Since inelastic Raman
scattering is sensitive to different symmetry charge modulations
(when polarizers are used on the incident and scattered light) it
can provide information about the symmetry of the CDW state and of the many-body charge excitations.  We expect our results should be
relevant to different experimental systems that display
charge-density-wave order via nesting on a bipartite lattice at
half filling, especially in compounds which are three-dimensional
like BaBiO$_3$ and Ba$_{1-x}$K$_{x}$BiO$_3$~\cite{cdw_exp1,cdw_exp2,cdw_exp3} because
DMFT is most accurate in higher spatial dimensions; it may also be
relevant to some layered two-dimensional systems, at least in a
semi-quantitative fashion. Our work is the next step in recent results on
transport, optical conductivity, and nonresonant inelastic X-ray
scattering in CDW systems~\cite{krishnamurthy, MSF1, MSF2} to the
realm of resonant inelastic light scattering. Since experimental inelastic
light scattering work on CDW systems was focused on the
Raman scattering of the soft phonon modes, future experimental work should examine the electronic scattering directly. Hence this work has the potential to
be directly relevant to the next generation of Raman experiments on strongly correlated CDW materials.

One of the simplest models which posseses static CDW ordering at low temperature is
the Falicov-Kimball model~\cite{falicov_kimball}. Historically,
this model was introduced in 1969 to describe metal-insulator
transitions in rare-earth compounds and transition-metal oxides.
Later, it was found that it has an exact solution within
DMFT~\cite{brandt_mielsch1} (for a review see
Ref.~\cite{freericks_review}). The Falicov-Kimball model has two
kinds of particles: itinerant electrons and localized electrons.
Mobile electrons hop from site to site with  a hopping integral
between nearest neighbors and they interact with the localized
electrons when both sit on the same site (the interaction energy
is $U$); we denote the itinerant electron creation (annihilation)
operator at site $i$ by $\hat d_i^\dagger$ ($\hat d_i^{}$) and the
local electron creation (annihilation) operator at site $i$ by
$\hat f_i^\dagger$ ($\hat f_i^{}$). The model has commensurate
(chessboard) CDW order at half filling and this is the main
property we exploit here. Brandt and Mielsch determined the
formalism for calculating the ordered-phase Green's
functions~\cite{brandt_mielsch2} shortly after Metzner and
Vollhardt introduced the idea of the many-body problem simplifying
in large dimensions~\cite{metzner_vollhardt}. The CDW order
parameter is known to display anomalous behavior at weak
coupling~\cite{vandongen,chen_freericks}, and higher-period
ordered phases exist on the Bethe
lattice~\cite{freericks_swiss}. 

\section{DMFT for the CDW phase of the Falicov-Kimball model}

As mentioned above, the Falicov-Kimball model possesses the
possibility for a transition into a commensurate CDW phase with
doubly modulated (chessboard-like) density of charge, when both
the itinerant and localized particles are half-filled. Since the hypercubic lattice is a bipartite lattice, implying
that it is separated into two sublattices (called $A$ and $B$)
with the nearest-neighbor hopping being nonzero only between the
different sublattices, the CDW order corresponds to the case
where the average filling of the electrons remains uniform on each
sublattice, but changes from one sublattice to another. We start by
writing the Falicov-Kimball model Hamiltonian as the sum of its
local and nonlocal parts
\begin{equation}\label{eq1}
  \hat{H}=\sum_{ia}\hat{H}_{i}^{a}-
  \sum_{ijab}t_{ij}^{ab}\hat{d}_{ia}^{\dag}\hat{d}_{jb}^{},
\end{equation}
where $i$ and $a=A$ or $B$ are the site and sublattice indices,
respectively, and $t_{ij}^{ab}$ is the hopping matrix, which is
nonzero only between different sublattices
($t_{ij}^{AA}=t_{ij}^{BB}=0$). The local Hamiltonian is equal to
\begin{equation}\label{eq2}
  \hat{H}_{i}^{a}=U\hat{n}_{id}^{a}\hat{n}_{if}^{a}-
  \mu_{d}^{a}\hat{n}_{id}^{a}-\mu_{f}^{a}\hat{n}_{if}^{a};
\end{equation}
with the number operators of the itinerant and localized electrons
given by $\hat n_{id}=\hat d_i^\dagger \hat d_i^{}$ and $\hat
n_{if}=\hat f_i^\dagger\hat f_i^{}$, respectively. Note that we
have introduced different chemical potentials for the different
sublattices. This is convenient for computations, because it
allows us to work with a fixed order parameter, rather than
iterating the DMFT equations to determine the order parameter.  That method of iterative solution is subject to critical slowing down near $T_c$, while working with a fixed order parameter is not.
The equilibrium solution occurs when the chemical
potential is uniform throughout the system ($\mu^A_d=\mu^B_d$ and
$\mu^A_f=\mu^B_f$), which is the unique condition used to find the order parameter at a given temperature.

We apply the DMFT, which provides an exact solution for the
Falicov-Kimball model in the limit of infinite spatial dimensions.
In contrast to the uniform case~\cite{sh_v_f_d1,sh_v_f_d2}, in the
CDW phase, the DMFT equations become matrix equations. Since the DMFT
solutions for the chessboard phase are described in detail in previous work~\cite{MSF1,MSF2}, we concentrate only
on a few basic points as a summary and to establish our notation. The first step of DMFT is to scale the hopping
matrix element as $t=t^*/2\sqrt{D}$~\cite{metzner_vollhardt} (we
use $t^*=1$ as the unit of energy) and then take the limit of
infinite dimensions $D\to\infty$. The self-energy is then local:
\begin{equation}\label{eq3}
  \Sigma_{ij}^{ab}(\omega)=\Sigma^{a}(\omega)\delta_{ij}\delta_{ab},
\end{equation}
and in the case of two sublattices has two values
$\Sigma^{A}(\omega)$ and $\Sigma^{B}(\omega)$. Now, we can write
the solution of the Dyson equation (in momentum space)
in a matrix form
\begin{equation}
\label{eq4}
  {\mathbf G}_{\bm k}(\omega)=\left[{\mathbf z}(\omega)-{\mathbf t}_{\bm k}\right]^{-1},
\end{equation}
where the irreducible part ${\mathbf z}(\omega)$ and hopping term
${\mathbf t}_{\bm k}$ are represented by the following $2\times2$
matrices:
\begin{eqnarray}\label{eq5}
  {\mathbf z}(\omega)&=\left ( \begin{array}{cccc}
  \omega+\mu^{A}_{d}-\Sigma^{A}(\omega) & 0  \\
  0 & \omega+\mu^{B}_{d}-\Sigma^{B}(\omega) \\
  \end{array}\right ),\quad
  {\mathbf t}_{\bm k }&=\left (\begin{array}{cccc}
  0 & \epsilon_{\bm k}  \\
  \epsilon_{\bm k} & 0 \\
  \end{array}\right ),
\end{eqnarray}
with $\epsilon_{\bm k}=t^*\lim\limits_{D\to\infty}\sum_{i=1}^D\cos ({k}_i)/\sqrt{D}$.

The second step of DMFT is to map the lattice Green's function
onto a local problem by means of the dynamical mean field. Since
there are two sublattices, a dynamical mean field
$\lambda^a(\omega)$ is introduced on each of them. As a result,
the local lattice Green's function on each sublattice becomes:
\begin{equation}\label{eq6}
  G^{aa}(\omega)=\frac{1}{\omega+\mu^a_d-\Sigma^a(\omega)-\lambda^a(\omega)}.
\end{equation}

The third equation that closes the system of equations for
$G^{aa}(\omega)$, $\Sigma^a(\omega)$ and $\lambda^a(\omega)$ is
obtained from the condition that the local Green's function is
defined as the Green's function of an impurity problem with the same
dynamical mean field $\lambda^{a}(\omega)$. Such a problem can be
exactly solved for the Falicov-Kimball model and the result is equal to
\begin{equation}\label{eq7}
  G^{aa}(\omega)=\frac{1-n_f^a}{\omega+\mu^a_d-\lambda^a(\omega)}
  + \frac{n_f^a}{\omega+\mu^a_d-U-\lambda^a(\omega)},
\end{equation}
where $n_f^a$ is an average concentration of the localized
electrons on sublattice $a$ which is found from the
equilibrium condition of a uniform chemical potential ($\mu_{f}^{A}-\mu_{f}^{B}=0$).

These equations are self-consistently solved numerically. In
Ref.~\cite{MSF1}, we analyzed the evolution of the DOS in
the CDW-ordered phase. We summarize the main points which are
needed here. At $T=0$, a real gap develops of magnitude $U$ with
square root singularities at the band edges. As the temperature
increases, the system develops substantial subgap DOS which are
thermally activated within the ordered phase until $T$ is raised high enough that the system enters the normal phase.  Plots of the DOS
can be found in Ref.~\cite{MSF1}. Note that the singular behavior
occurs for one of the ``inner'' band edges on each sublattice, and
that the subgap states develop very rapidly as the temperature
rises. Furthermore, the DOS on each sublattice is related to the DOS on the other sublattice by a reflection about $\omega=0$.

\section{Inelastic light scattering}

For an electronic system with nearest-neighbor hopping, the
interaction with a weak external transverse electromagnetic field determined by the vector potential
${\bm A}$ is described by the
Hamiltonian~\cite{shastry_shraiman1,shastry_shraiman2}:
\begin{equation}\label{eq8}
  H_{\textrm{int}}=-\frac{e}{\hbar c}
  \sum_{\bm k}{ {\bm j}(\bm k)\cdot \bm A(-\bm k)}
  +\frac{e^{2}}{2\hbar^{2}c^{2}}
  \sum_{ {\bm k \bm k'}}{A_{\alpha}(-\bm k)\gamma_{\alpha,\beta}(\bm k+ \bm k')A_{\beta}(-
  \bm k')},
\end{equation}
where the number current operator and stress tensor for itinerant
electrons are equal to
\begin{equation}\label{eq9}
  j_{\alpha}(\bm q)=\sum_{ab \bm k}
  \frac{\partial t_{ab}(\bm k)}
  {\partial k_{\alpha}}\hat{d}_{a}^{\dag}(\bm k+\bm q/2)\hat{d}_{b}(\bm k-{\bm q}/2)
  \quad \textrm{and} \quad
  \gamma_{\alpha,\beta}(\bm q)=\sum_{ab\bm k}
  \frac{\partial^{2} t_{ab}(\bm k)}{\partial k_{\alpha}
  \partial k_{\beta}}\hat{d}_{a}^{\dag}(\bm k+\bm q/2)\hat{d}_{b}(\bm k-\bm q/2),
\end{equation}
respectively. Here $t_{ab}(\bm k)$ are the components of the
$2\times 2$ hopping matrix in Eq.~(\ref{eq5}). The formula for the
inelastic light scattering cross section derived by Shastry and
Shraiman is equal to~\cite{shastry_shraiman1,shastry_shraiman2}
\begin{equation}\label{eq10}
  R(\bm q,\Omega)=2\pi\sum_{i,f}
  \frac{e^{-\beta\varepsilon_{i}}}{Z}
  \delta(\varepsilon_{f}-\varepsilon_{i}-\Omega)
  \left|g(\bm k_{i})g(\bm k_{f})e_{\alpha}^{i}e_{\beta}^{f}
  \left\langle f\left|\hat{M}^{\alpha\beta}(\bm q)\right|i\right\rangle \right|^{2}.
\end{equation}
It describes the scattering of band electrons by photons with
$\Omega=\omega_{i}-\omega_{f}$ and $\bm q=\bm k_{i}-\bm k_{f}$
being the transferred energy and momentum, respectively, $\bm
e^{i(f)}$ is the polarization of the initial (final) states of the
photons and $\varepsilon_{i(f)}$ denotes the electronic
eigenstates. The quantity $g(\bm q)=(hc^{2}/V\omega_{\bm
q})^{1/2}$ is called the ``scattering strength'' with $\omega_{\bm
q}=c|\bm q|$, and $Z={\rm Tr}~\exp(-\beta \hat H)$ the partition function. The scattering
operator $\hat{M}(\bm q)$ is constructed from both the number
current operator and the stress tensor; it has both nonresonant
and resonant contributions
\begin{equation}\label{eq11}
  \left\langle f\left|\hat{M}^{\alpha\beta}(\bm q)\right|i\right\rangle
  =\left\langle f\left|\gamma_{\alpha,\beta}(\bm q)\right|i\right\rangle
  +\sum_{l}\Biggl(\frac{\left\langle f\left|j_{\beta}(\bm k_{f})
  \right|l\right\rangle  \left\langle l\left|j_{\alpha}(-\bm k_{i})
  \right|i\right\rangle }{\varepsilon_{l}-\varepsilon_{i}-\omega_{i}}
  +\frac{\left\langle f\left|j_{\alpha}(-\bm k_{i})
  \right|l\right\rangle  \left\langle l\left|j_{\beta}(\bm k_{f})
  \right|i\right\rangle }{\varepsilon_{l}-\varepsilon_{i}+\omega_{f}}
  \Biggr)
\end{equation}
with the sum $l$ over intermediate states. After substituting
into the cross section formula, one obtains three terms in the cross section: a
nonresonant term; a mixed term; and a pure resonant term (because
it is constructed from the square of the scattering operator). The
components of the cross section can be extracted from the
appropriate correlation functions (response functions) first
calculated on the imaginary Matsubara frequencies and then analytically continued onto the real
axis~\cite{sh_v_f_d2}. Hence, we concentrate on the
light-scattering response function $\chi(\bm q,\Omega)$, which is
related to the cross section and has nonresonant, mixed and
resonant contributions:
\begin{equation}\label{eq12}
    R(\bm q,\Omega)=\frac{2\pi g^{2}(\bm k_{i})g^{2}(\bm k_{f})}{1-\exp(-\beta \Omega)}
    \chi(\bm q,\Omega),\quad \chi(\bm q,\Omega)=\chi_{N}(\bm q,\Omega)+\chi_{M}(\bm q,\Omega)+\chi_{R}(\bm q,\Omega).
\end{equation}

Inelastic light scattering examines charge excitations of
different symmetries by employing polarizers on both the incident
and scattered light. The $A_{\rm 1g}$ symmetry has the full
symmetry of the lattice and is primarily measured by taking the
initial and final polarizations to be $\bm e^{i}=\bm
e^{f}=(1,1,1,1,\ldots)$. The $B_{\rm 1g}$ symmetry involves
crossed polarizers: $\bm e^{i}=(1,1,1,1,\ldots)$ and $\bm
e^{f}=(-1,1,-1,1,\ldots)$; while the $B_{\rm 2g}$ symmetry is
rotated by 45 degrees; it requires the polarization vectors to
satisfy $\bm e^{i}=(1,0,1,0,\ldots)$ and $\bm
e^{f}=(0,1,0,1,\ldots)$. For Raman scattering  ($\bm q=0$), it is
easy to show that for a system with only nearest-neighbor hopping
and in the limit of large spatial dimensions, the $A_{\rm 1g}$
sector has contributions from nonresonant, mixed and resonant
scattering, the $B_{\rm 1g}$ sector has contributions from
nonresonant and resonant scattering only, and the $B_{\rm 2g}$
sector is purely resonant~\cite{freericks_deveraux1,freericks_deveraux2}.

An analysis of the total electronic Raman spectra for the uniform phase of the
Falicov-Kimball model has already been completed~\cite{sh_v_f_d2}.
A full calculation of the nonresonant inelastic light scattering for all $\bm q$  in the CDW chess-board phase has also been presented~\cite{MSF2}. Here we focus on the total
response including the mixed and resonant contributions for the
Raman scattering ($\bm q=0$) in CDW phase.

\section{Mixed and resonant responses}

The way we determine the mixed and resonant response functions is
as follows: we construct the corresponding multi-time correlation
function in terms of the generalized polarizations, then perform
a Fourier transformation to the imaginary Matsubara frequencies and
finally analytically continue onto the real frequency axis to
extract the response function. Such a procedure requires a lot of
algebra and analysis~\cite{sh_v_f_d2}; we do not present all of the details here, but instead we summarize the main points.

\begin{figure}[tb]\label{fig:diagr_mix}
\noindent\includegraphics[scale=.8]{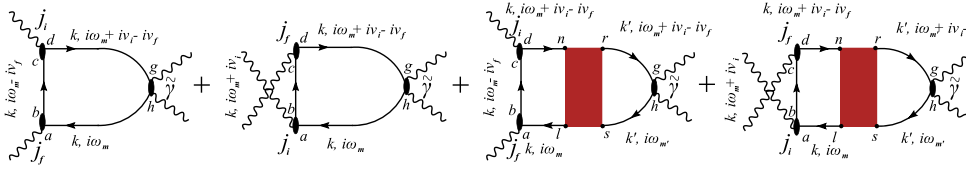}
   \caption{Feynman diagrams for the generalized polarizations of the mixed response function.
   Due to the properties of the dynamic irreducible charge vertex of the Falicov-Kimball model, we have
   $m=m^\prime$.}
\end{figure}

The mixed response function is extracted from the multi-time
correlation function which is built on three operators: one stress
tensor and two current operators, as follows
\begin{equation}\label{eq13}
 \chi_{\tilde{\gamma},\tilde{f},\tilde{i}}(\tau_{1},\tau_{2},\tau_{3})=\left\langle
   T_{\tau}\tilde{\gamma}(\tau_{1})\tilde{j}^{(f)}(\tau_{2})\tilde{j}^{(i)}(\tau_{3})\right\rangle .
\end{equation}
The symbol $T_{\tau}$ is a time ordering operator and the tilde
denotes contractions with the polarization vectors
[$\tilde{\gamma}=\sum\limits_{\alpha\beta}e_{\alpha}^{i}\gamma_{\alpha,\beta}(\bm q)e_{\beta}^{f}$ and $\tilde{j}^{(i,f)}=\sum\limits_{\alpha} e_{\alpha}^{i,f} j_{\alpha}(\mp\bm k_{i,f})$].
Furthermore, we perform the Fourier transformation from the imaginary
times $\tau_{1},\tau_{2}$, and $\tau_{3}$ to the imaginary
Matsubara frequencies $i\nu_{i}, i\nu_{f}$, and
$i\nu_{i}-i\nu_{f}$. As a result, the correlation function is
represented as a sum over Matsubara frequencies of the generalized
polarizations $\Pi_{m+i,m+i-f,m}$:
\begin{equation}\label{eq14}
  \chi_{\tilde{\gamma},f,i}(i\nu_i-i\nu_f,i\nu_f,-i\nu_i)
  =T\sum\limits_{m}\left [\Pi_{m-f,m+i-f,m}+\Pi_{m+i,m+i-f,m}\right
  ],
\end{equation}
where we introduced the shorthand notation
$\Pi_{m-f,m+i-f,m}=\Pi(i\omega_{m}-i\nu_{f},i\omega_{m}+i\nu_{i}-i\nu_{f},i\omega_{m})$
for the dependence on the fermionic $i\omega_{m}=i\pi T(2m+1)$ and
bosonic $i\nu_{l}=i2\pi Tl$ Matsubara frequencies. In the case of
a CDW ordered phase, the Feynman diagrams for the generalized
polarizations $\Pi_{m,m+l}$ are shown in Fig.~\ref{fig:diagr_mix},
where we introduce additional sublattice indexes $a$ to $s$.

Now one has to carefully perform the analytic continuation to the
real axis ($i\nu_{i,f}\to\omega_{i,f}\pm i0^{+},
i\nu_{i}-i\nu_{f}\to\Omega\pm i0^{+}$) and replace the sum over
Matsubara frequencies by an integral over the real axis. Then the
mixed response function is expressed directly in terms of the
generalized polarizations
\begin{align}\label{eq15}
  \chi_{M}(\bm q,\Omega)
  &=\frac{1}{(2\pi i)^{2}}\int\limits_{-\infty}^{+\infty} d \omega
  \left[f(\omega)-f(\omega+\Omega)\right]
  \nonumber\\
  &\times\textrm{Re}\Bigl\{\Pi(\omega-\omega_{f}+i0^{+},\omega+\Omega+i0^{+},\omega-i0^{+})
  -\Pi(\omega-\omega_{f}+i0^{+},\omega+\Omega-i0^{+},\omega-i0^{+})
  \nonumber \\
  &+\Pi(\omega-\omega_{f}-i0^{+},\omega+\Omega+i0^{+},\omega-i0^{+})
  -\Pi(\omega-\omega_{f}-i0^{+},\omega+\Omega-i0^{+},\omega-i0^{+})
  \\
  &+\Pi(\omega+\omega_{i}+i0^{+},\omega+\Omega+i0^{+},\omega-i0^{+})
  -\Pi(\omega+\omega_{f}+i0^{+},\omega+\Omega-i0^{+},\omega-i0^{+})
  \nonumber \\
  &+\Pi(\omega+\omega_{f}-i0^{+},\omega+\Omega+i0^{+},\omega-i0^{+})
  -\Pi(\omega+\omega_{f}-i0^{+},\omega+\Omega-i0^{+},\omega-i0^{+})
  \Bigr\},\nonumber
\end{align}
where $f(\omega)=1\left/[\exp(\beta\omega)+1]\right.$ is the Fermi-Dirac
distribution function.

The mixed contribution is nonzero only for the $A_{\rm 1g}$
symmetry; for other symmetries it vanishes after the summation
over wave vectors. The next step is to calculate these generalized
polarizations. There are two types of diagrams for the generalized
polarizations (see Fig.~\ref{fig:diagr_mix}): the bare loop and
the renormalized loop. The reducible charge vertex $\tilde\Gamma^{ab}$
(shaded rectangle in Fig.~\ref{fig:diagr_mix}) is defined from the
Bethe-Salpeter-like equation through the irreducible one
$\Gamma_{a}$ which is local in the DMFT approach on each
sublattice~\cite{MSF2} and has the same functional form as in the uniform
phase~\cite{charge_vertex1,charge_vertex2,charge_vertex3}. Also, we used the fact that the total
reducible charge vertex is a diagonal function of frequencies for
the Falicov-Kimball model~\cite{charge_vertex1,charge_vertex2,charge_vertex3}; for other models,
where the vertex is no longer diagonal, the analysis is
more complicated. The final expression for the generalized
polarization is too cumbersome to be presented here. 

\begin{figure}[tb]\label{fig:diagr_res}
\noindent \includegraphics[scale=.9]{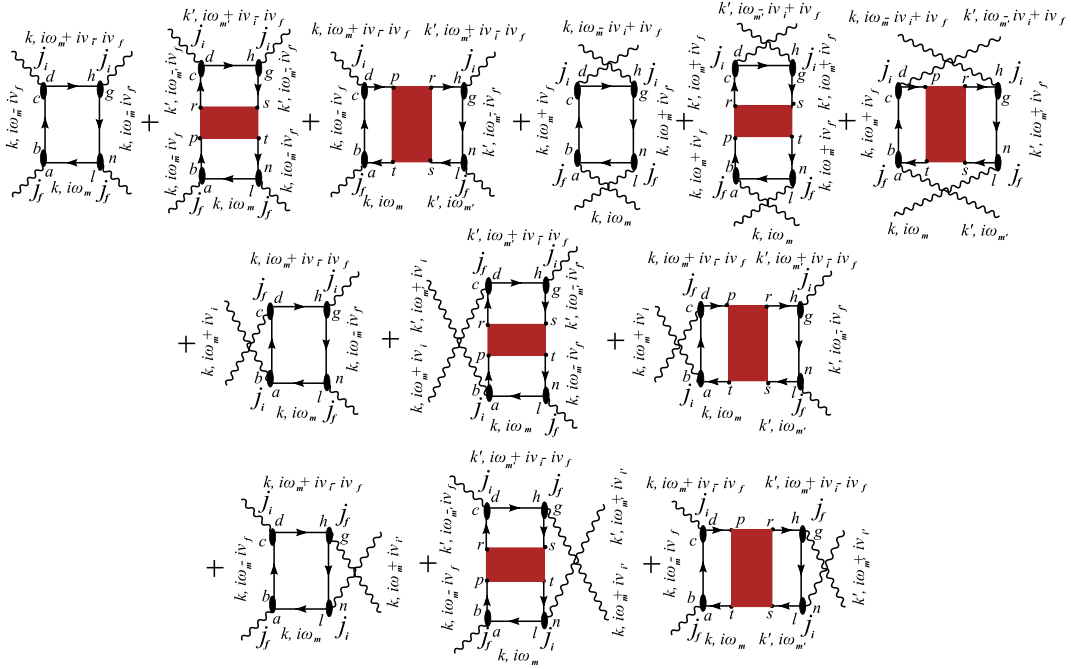}
   \caption{Feynman diagrams for the generalized polarizations of the resonant response function.
   Both the renormalized and bare loop diagrams contribute in all symmetries ($A_{1g}$, $B_{1g}$ and $B_{2g}$).}
\end{figure}

Similar to the mixed response, the resonant response function is
constructed from a multi-time correlation function which is
built on four current operators
\begin{equation}\label{eq16}
 \chi_{i,f,f,i}(\tau_{1},\tau_{2},\tau_{3},\tau_{4})=\left\langle
   T_{\tau}(j^{(i)}(\tau_{1})j^{(f)}(\tau_{2})j^{(f)}(\tau_{3})j^{(i)}(\tau_{4})\right\rangle.
\end{equation}
Furthermore, we perform the same formal analytic continuation procedure as for the mixed response
function. In Fig.~\ref{fig:diagr_res}, we present 
the Feynman diagrams for the generalized polarizations which
contribute to the resonant response function.

For the resonant response function, the analytical continuation
onto the real axis is quite complex, but the general method remains the
same and the final formula for the resonant response function is
similar to the mixed response function in  Eq.~(\ref{eq15}) (see
Ref.~\cite{sh_v_f_d2} for results in the normal phase). In contrast to the nonresonant and mixed
response, the resonant response contributes to all symmetries.
In addition, both the bare and renormalized loops in
Fig.~\ref{fig:diagr_res} are present in the resonant response function.
In the $B_{\rm 2g}$ symmetry we have only the
resonant response, in the $B_{\rm 1g}$ symmetry we have both
nonresonant and resonant responses, and in the $A_{1g}$ symmetry
we have all three responses.

\section{Results and conclusions}

\begin{figure}[tb]\label{fig:dos_resp}
\noindent\includegraphics[scale=.8]{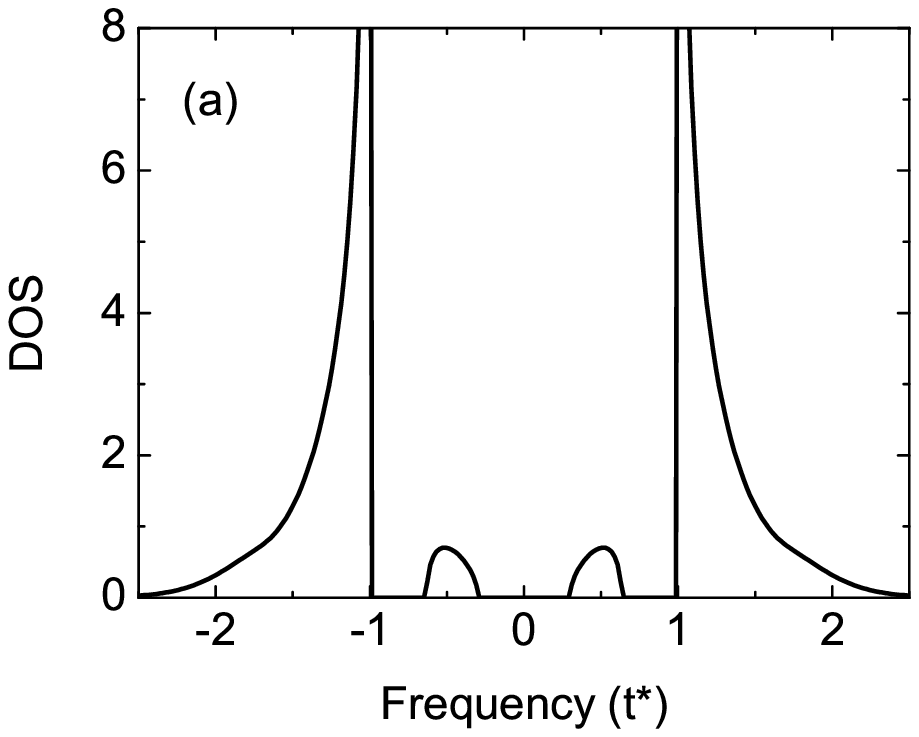} \qquad
\includegraphics[scale=.8]{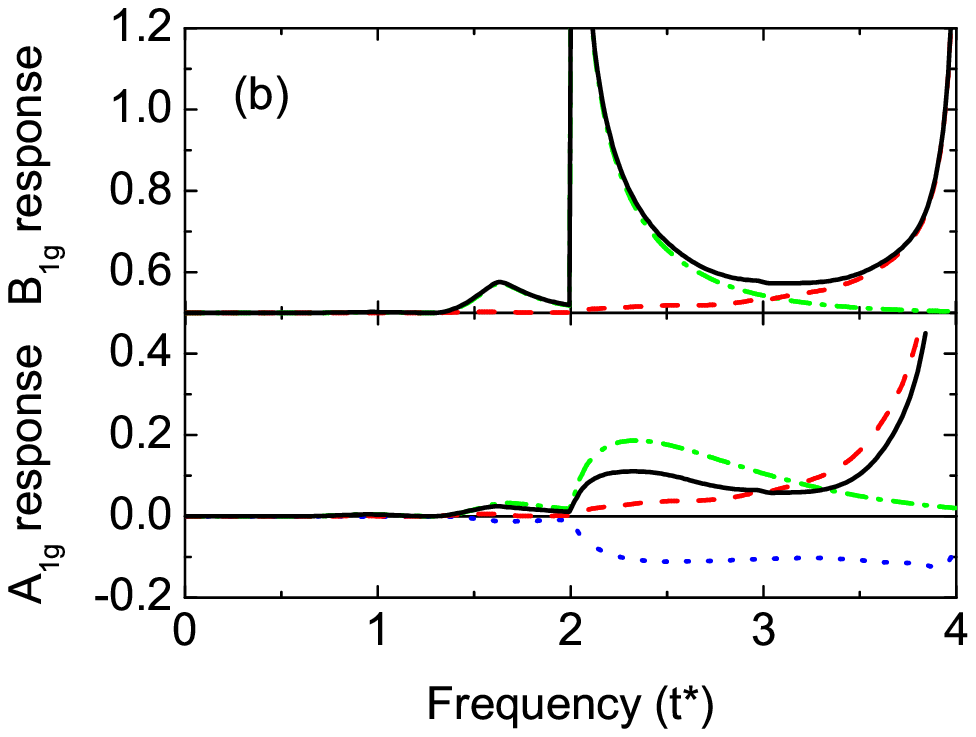}
   \caption{(a) DOS and (b) different contributions to the Raman
   responses ($\omega_i=4$) at $T=0.05$ and $U=2$.
   The solid line corresponds to the total response,
   the dashed-dotted line corresponds to the nonresonant contribution,
   the dashed line corresponds to the resonant contribution,
   and
   the dotted line corresponds to the mixed contribution. }
\end{figure}

A detailed analysis of the single particle DOS in
the chessboard CDW phase of the Falicov-Kimball model has
been presented earlier~\cite{MSF1}. Here we show
results for Raman scattering in the CDW phase with
$U=2$  and temperature $T=0.05$
(which lies below the critical temperature $T_{c}\approx 0.0769$). In
Fig.~\ref{fig:dos_resp}, we show the DOS (left panel) and the
different contributions to the Raman response for the $A_{\rm 1g}$ and
$B_{\rm 1g}$ symmetries.
\begin{figure}[tb]
\noindent\includegraphics[scale=.8]{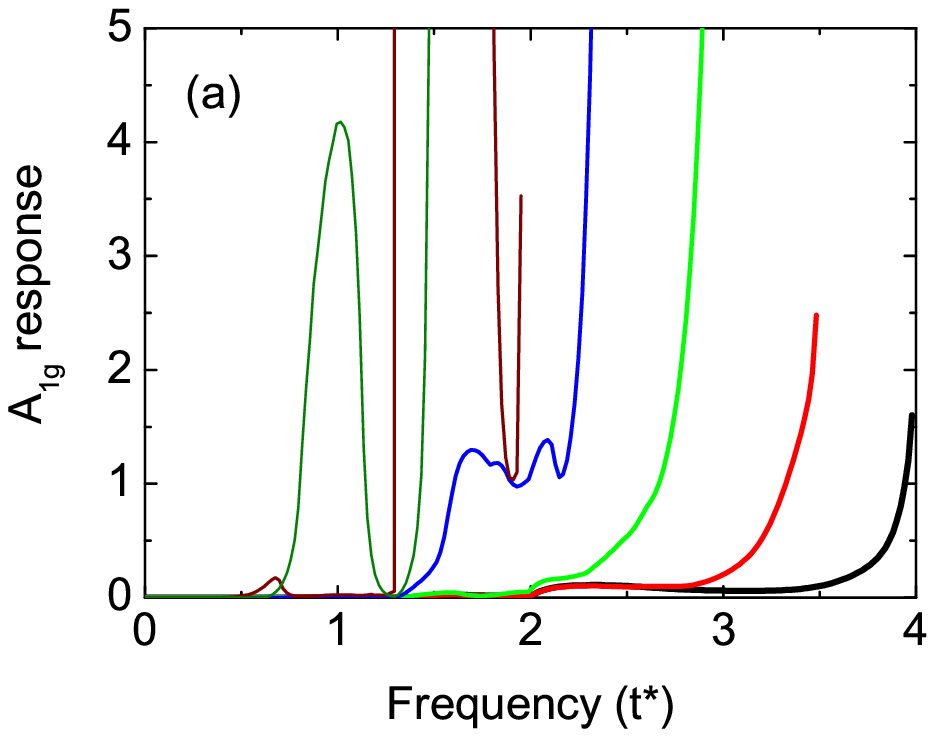} \qquad
\includegraphics[scale=.8]{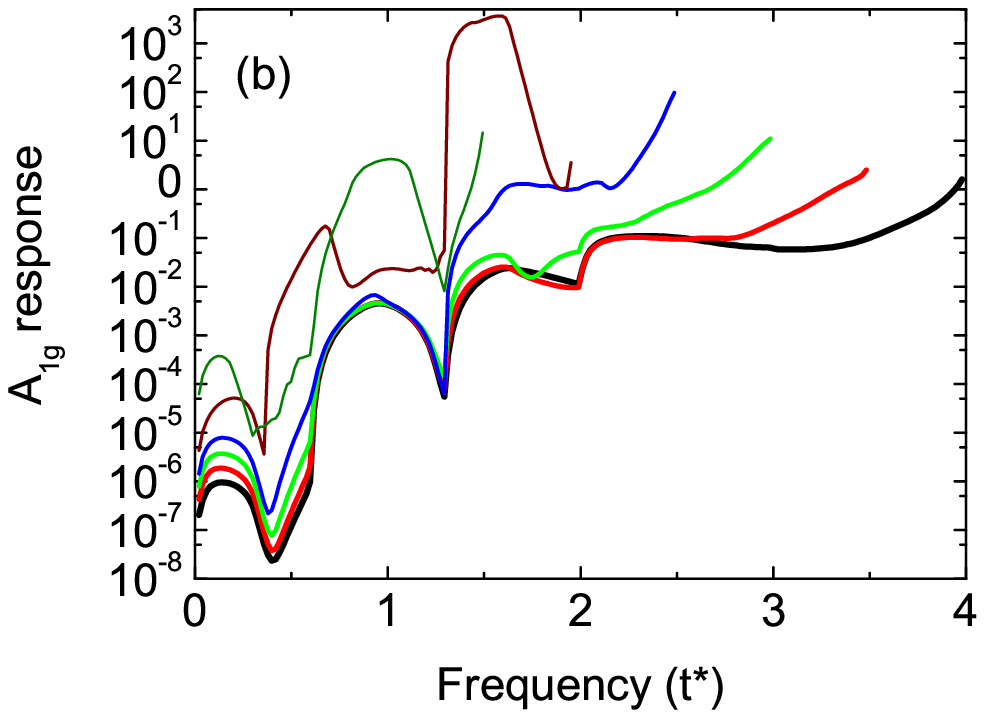}
   \caption{Total Raman response for the $A_{\rm 1g}$ symmetry at $T=0.05$ for $U=2$ on (a) a linear
   and (b) a logarithmic scale. Different curves correspond to different incident photon
   energies ranging from $\omega_i=1.5$ to $\omega_i=4.0$ in steps of $0.5$.}\label{fig:a1g}
\end{figure}
In Fig.~\ref{fig:dos_resp}~(a) one can see the main
features of the DOS for chessboard phase at intermediate $T$: the gap of width $U$
edged by singularities at $\omega=\pm U/2$ and partially filled by
subgap states placed at $\omega=\pm E/2$ ($E\approx 1$ for the case of $U=2$ and $T=0.05$). As a result, the
scattering spectra displays features (peaks) at frequencies
$\Omega=(U-E)/2$, $E$, $(U+E)/2$, and $U$.
Such features (peaks) were already observed for the optical
conductivity~\cite{MSF1} and for the nonresonant Raman and X-ray
responses~\cite{MSF2}. In addition, there can also be features at
$\omega_{i}-U$, $\omega_{i}-(U+E)/2$, $\omega_{i}-E$, and
$\omega_{i}-(U-E)/2$.

In the case of $B_{\rm 1g}$ symmetry [Fig.~\ref{fig:dos_resp}~(b)], there is a large peak in the
nonresonant response at a frequency equal to $\Omega=U$ that
reflects the transitions between the states above and below the
gap. The peak has a square root-like singularity, that comes
from the shape of DOS when there is CDW order and the fact that there is no screening for the nonresonant response in this symmetry channel. The smaller peak
corresponds to transitions between states of the upper
(bottom) band and the lower (upper) subgap states at
$[\Omega=(U+E)/2]$. In addition to the nonresonant response, the
resonant one has contributions from the renormalized charge
excitations (Fig.~\ref{fig:diagr_res}) which reduce the
high-energy transitions. As was mentioned above, all contributions
contribute in the total response for the $A_{\rm 1g}$ symmetry [Fig.~\ref{fig:dos_resp}~(b)]. Because of
the charge renormalization in the nonresonant, mixed and resonant
response functions, the total response is smaller than in the $B_{\rm 1g}$
symmetry and there is no square root singularity.

\begin{figure}[tb]
\noindent\includegraphics[scale=.8]{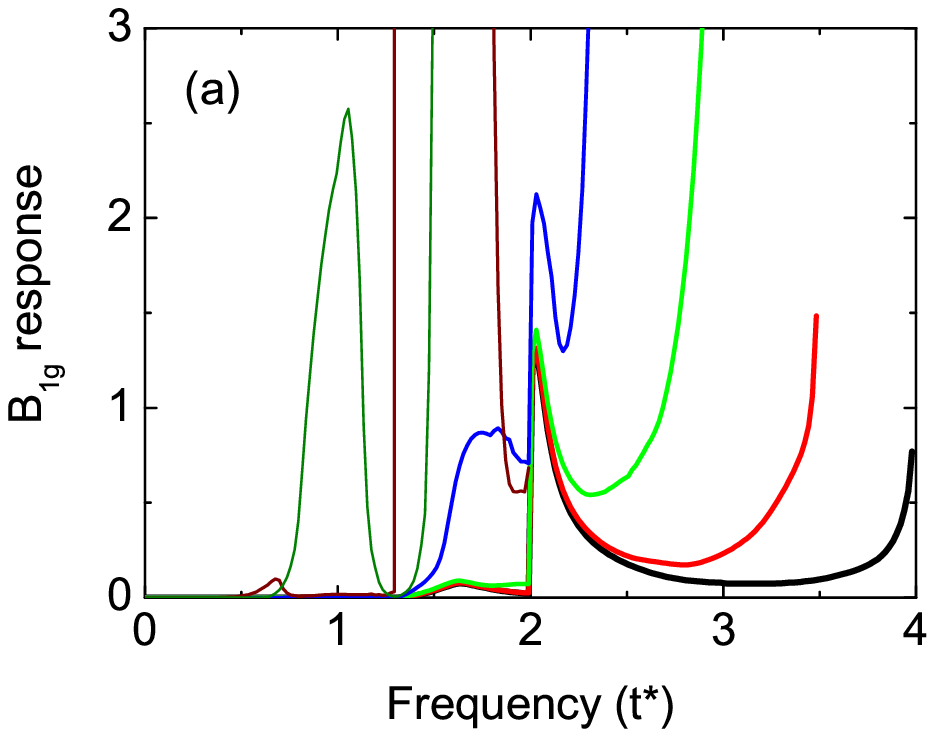} \qquad
\includegraphics[scale=.8]{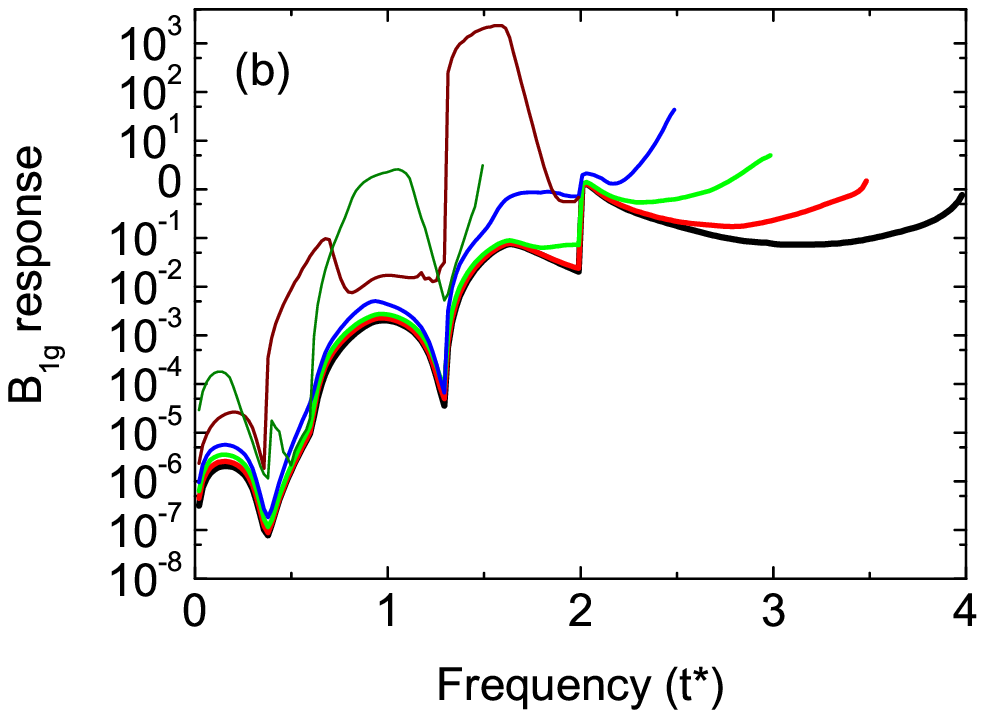}
   \caption{Total Raman response for the $B_{\rm 1g}$ symmetry at $T=0.05$ for $U=2$ on (a) a linear
   and (b) a logarithmic scale. Different curves correspond to different incident photon
   energies ranging from $\omega_i=1.5$ to $\omega_i=4.0$ in steps of $0.5$.}\label{fig:b1g}
\end{figure}
\begin{figure}[tb]
\noindent\includegraphics[scale=.8]{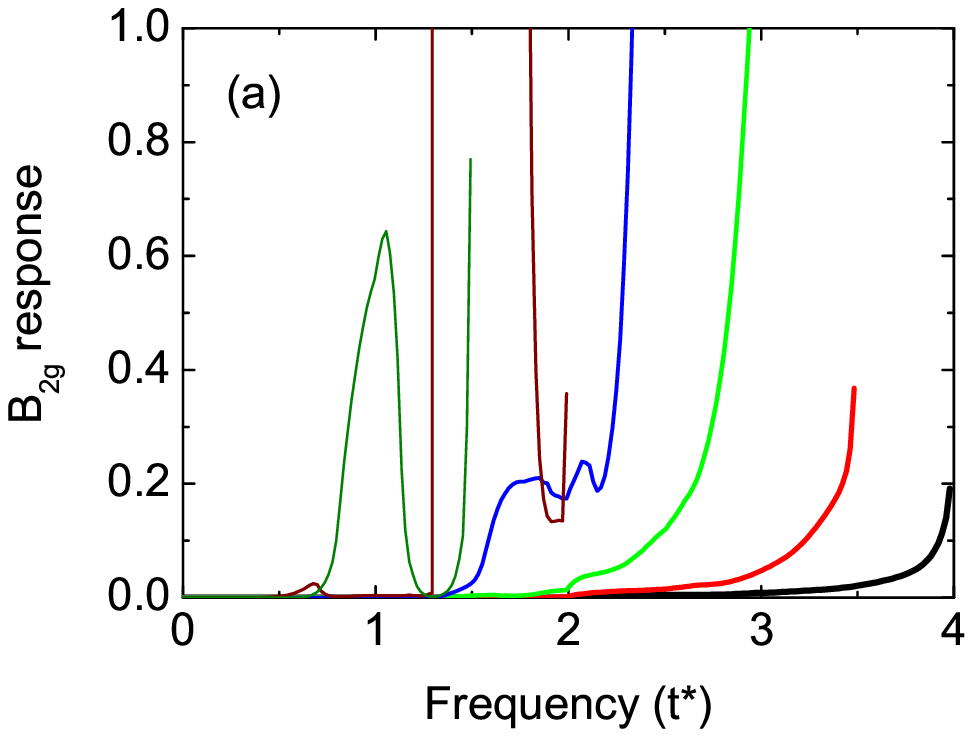} \qquad
\includegraphics[scale=.8]{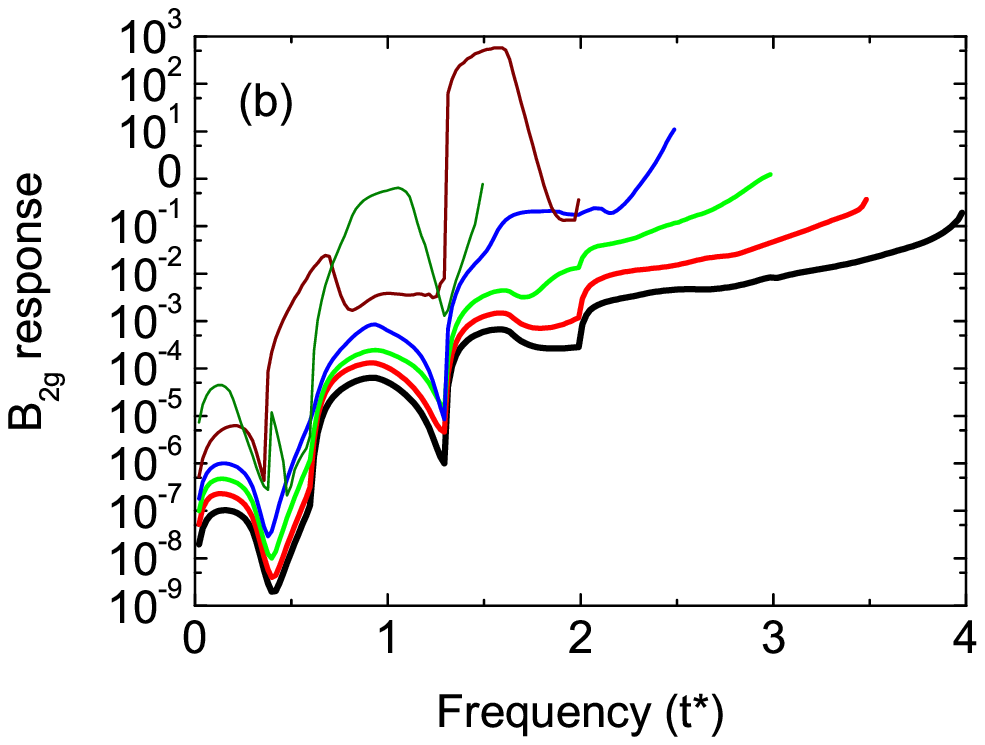}
   \caption{Total Raman response for the $B_{\rm 2g}$ symmetry at $T=0.05$ for $U=2$ on  (a) a linear
   and (b) a logarithmic scale. Different curves correspond to different incident photon
   energies ranging from $\omega_i=1.5$ to $\omega_i=4.0$ in steps of $0.5$.}\label{fig:b2g}
\end{figure}

The total Raman response is presented in set of
Figs.~\ref{fig:a1g}--\ref{fig:b2g} for the different symmetry channels
and for different energies of the incident photons. One can see the
main features which were already observed for the optical
conductivity and nonresonant scattering: four peaks at
$\Omega=(U-E)/2,\hspace{0.1cm}E,\hspace{0.1cm}(U+E)/2$, and $U$
which correspond to the different interband transitions. In addition, the resonant
and mixed contributions strongly modify the nonresonant response:
there is a strong enhancement of the scattering when the energy of the incident
photon is close to the energy of the interband transitions and there is the appearance of
additional features (peaks) at the frequencies $\Omega=\omega_{i}-U$,
$\omega_{i}-(U-E)/2$, $\omega_{i}-E$, and $\omega_{i}-(U+E)/2$
as measured from the energy of the incident photon.

The resonant response is particularly large near the transferred energy $\Omega\approx 1.5$ when $\omega_i\approx 2$.  By examining results on a finer grid of photon frequencies (not shown here) we establish that the resonant profile has a narrow full width at half max of much less than 0.1, and a very sharp dependence on $\omega_i$ (the peak height drops by more than three orders of magnitude by the time $\omega_i=2.1$ or 1.9). In addition, the resonant response in this region does not depend too strongly on the symmetry channel of the scattering.  Note how similar the curves appear (on a log scale) for the different symmetry channels, especially for transferred energies away from $U$ where the gap edge creates sharp features in the $B_{\rm 1g}$ channel. Finally, there are joint resonances, as the lower energy peaks do resonate with the large peak, especially for $\omega_i\approx 2$.  Similar resonant effects can be seen for the lower-energy peaks when the incident photon frequency is lower, but they are not as dramatic as what happens for the peak near $\Omega\approx 1.5$. These resonant effects could be strong signatures of the CDW phase in real materials.  The resonant enhancements in the normal phase do not produce such enormous peaks or have such sharp dependences on the incident photon frequencies; this behavior is arising predominantly from the CDW order.

In conclusion, we have examined the total electronic Raman scattering response for the spinless Falicov-Kimball model in the ordered CDW phase.  Space limitations allowed us to only consider one value of the interaction and temperature, but we see some interesting results, primarily the appearance of a huge resonantly enhanced peak near $\omega_i=U$ that is essentially independent of the symmetry channel.  Such a peak could be an important signal for experiments on these systems as indicating the appearance of the CDW phase via a direct measurment of the electronic charge dynamics. Future work will elaborate on how these features evolve with $U$ and $T$ and will provide additional details of the formalism that could not be included here.

\begin{theacknowledgments}
J.~K.~F.~was supported by the 
Department of Energy, Office of Basic Energy Sciences, under grant number DE-FG02-08ER46542. The collaboration was 
supported by the Computational Materials Science Network (CMSN) program of the Division of Materials Science and Engineering, Basic Energy Sciences, 
U.~S.~Department of Energy under grant number DE-FG02-08ER46540.
\end{theacknowledgments}

\end{document}